\documentclass
[prl,twocolumn,10pt,a4paper,superscriptaddress,showpacs]{revtex4}%
\usepackage{amsmath,graphicx}
\usepackage{amsmath}
\usepackage{graphicx}%
\usepackage{amsfonts}%
\usepackage{amssymb}
\def\opone{\leavevmode\hbox{\small1\kern-3.8pt\normalsize1}}

\begin{document}

\title{Experimental demonstration of an efficient quantum phase-covariant cloning and
its possible applications to simulating eavesdropping in quantum cryptography}
\author{Jiangfeng Du}
\email{djf@ustc.edu.cn}
\affiliation{Structural Research Laboratory and Department of Modern Physics, University
of Science and Technology of China, Hefei, 230027, P.R. China}
\affiliation{Department of Physics, National University of Singapore, 2 Science Drive 3,
Singapore 117542}
\affiliation{Centre for Quantum Computation, DAMTP, University of Cambridge, Wilberforce
Road, Cambridge CB3 0WA U.K.}
\author{Thomas Durt}
\affiliation{TONA-TENA Free University of Brussels, Pleinlaan 2, B-1050 Brussels, Belgium.}
\author{Ping Zou}
\affiliation{Structural Research Laboratory and Department of Modern Physics, University
of Science and Technology of China, Hefei, 230027, P.R. China}
\author{L.C. Kwek}
\affiliation{Department of Natural Sciences, National Institute of Education, Nanyang
Technological University, 1 Nanyang Walk, Singapore 637616}
\author{C.H. Lai}
\affiliation{Department of Physics, National University of Singapore, 2 Science Drive 3,
Singapore 117542}
\author{C.H. Oh}
\affiliation{Department of Physics, National University of Singapore, 2 Science Drive 3,
Singapore 117542}
\author{Artur Ekert}
\affiliation{Centre for Quantum Computation, DAMTP, University of Cambridge, Wilberforce
Road, Cambridge CB3 0WA U.K.}
\affiliation{Department of Physics, National University of Singapore, 2 Science Drive 3,
Singapore 117542}

\begin{abstract}
We describe a nuclear magnetic resonance (NMR) experiment which implements an
efficient one-to-two qubit phase-covariant cloning machine(QPCCM). In the
experiment we have achieved remarkably high fidelities of cloning, $0.848 $
and $0.844$ respectively for the original and the blank qubit. This
experimental value is close to the optimal theoretical value of $0.854$. We
have also demonstrated how to use our phase-covariant cloning machine for
quantum simulations of bit by bit eavesdropping in the four-state quantum key
distribution protocol.

\end{abstract}
\maketitle

The ``no-cloning" result~\cite{Wootters, Ghirardi} asserts that due to the
linearity of quantum mechanics unknown quantum states cannot be copied
perfectly. This notwithstanding one can design approximate quantum cloning
machines and address their optimality. The most notable example is the
universal quantum cloner (UQC) proposed by Bu\v{z}ek and Hillery~\cite{buzek}.
It has been studied in great details~\cite{UQCM} and a number of experimental
implementations of a $1\rightarrow2$ qubit UQC have been
proposed~\cite{Cummins,Antia,Fasel,huang}. Another important example is the
optimal quantum phase-covariant cloning machine (QPCCM)~\cite{FGGNP,NG,bruss}.
Unlike the UQC, it clones only subsets of states for which we have some
\textit{a priori} information. In the special case of the QPCCM operating on
qubits it has been shown that a class of states $\left\vert \psi\right\rangle
=\frac{1}{\sqrt{2}} \left(  \left\vert 0\right\rangle +e^{i\varphi}\left\vert
1\right\rangle \right)  $, called equatorial states, can be cloned up to the
fidelity $0.854$. As expected this value is slightly higher than the optimal
fidelity of the UQC ($0.833$). This is because even partial information about
the original state allows to optimize the cloning process and to obtain higher
fidelities of the clones. The phase-covariant cloners, which are the subject
of this paper, are of significant importance in quantum cryptography as they
provide the optimal eavesdropping technique for a large class of attacks on
the four state protocol (BB84)~\cite{BB84}. The properties of the QPCCM have
been extensively studied from the theoretical perspective~\cite{QPCCM,thomas},
however, on the experimental side, apart from an interesting recent optical
proposal by Fiurasek~\cite{Jaromir}, no actual realization of the QPCCM has
been reported.

Here we describe the first experimental implementation of the QPCCM. We use
the NMR technology to implement a modified two qubit network originally
designed by Niu and Griffiths~\cite{NG} (see Fig. 1). The simplicity of the
network allows to reduce the effects of decoherence and to obtain remarkably
high fidelities of the clones.

\begin{figure}[ptb]
\includegraphics[width=0.5 \columnwidth]{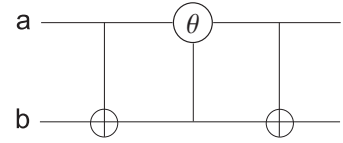}\caption{Quantum network of
the efficient phase-covariant cloning. It consists of two controlled-NOT gates
together with one controlled-rotation gate, where $R(\theta)=e^{-i\theta
\sigma_{y}/2}$ is a rotation by an angle $\theta$\ about the $y$\ axis in
Bloch-sphere. The upper and the lower horizontal lines correspond to the
original and the blank qubits respectively.}%
\end{figure}

This is in contrast with earlier approaches which were based on more
complicated three qubit networks~\cite{FGGNP}. If the complexity of related
three qubits experiments is of any guidance here, e.g. the NMR implementation
of the UQC~\cite{Cummins} , then substantial losses due to inhomogeneities of
the magnetic field and decoherence cannot be avoided with the current state of
the art technology. This, together with the stringent precision requirements,
lowers the fidelity (to about inconclusive $58\%$ in~\cite{Cummins}). A three
qubit network for for the $1\rightarrow2$ QPCCM, for example the one proposed
by Fuchs et al.~\cite{FGGNP}, would face similar problems.

Our version of the network is shown in Fig. 1. The net unitary operator has
the form
\[
U\left(  \theta\right)  =\left(
\begin{array}
[c]{cccc}%
1 & 0 & 0 & 0\\
0 & \cos\theta & \sin\theta & 0\\
0 & -\sin\theta & \cos\theta & 0\\
0 & 0 & 0 & 1
\end{array}
\right)  .
\]
When $\theta=\frac{\pi}{4}$, this unitary transform defined as $U^{opt}$
corresponds to an efficient optimal QPCCM. In fact, $U^{opt}$\ is just a
2-qubit square root of SWAP gate. Consider now an equatorial state of $a$
qubit, i.e., a state with a definite spin in the direction $n=\left(
\cos\varphi,\sin\varphi,0\right)  $. This state has the form $\left\vert
n\right\rangle =\frac{1}{\sqrt{2}}\left(  \left\vert 0\right\rangle
+e^{i\varphi}\left\vert 1\right\rangle \right)  $, whereas the $b$ qubit is in
the state $\left\vert 0\right\rangle $. $U^{opt}$\ transforms the input state
to $\rho_{ab}^{out}=U^{opt}\left\vert n\right\rangle \left\vert 0\right\rangle
\left\langle 0\right\vert \left\langle n\right\vert U^{opt+}$. The reduced
density matrix of two copies can be calculated (by tracing out another qubit)
as%
\[
\rho_{a}^{out}=\rho_{b}^{out}=\left(
\begin{array}
[c]{cc}%
3/4 & \sqrt{2}e^{-i\varphi}/4\\
\sqrt{2}e^{i\varphi}/4 & 1/4
\end{array}
\right)  .
\]
Note that the two copies are in fact symmetric. We use the fidelity
$F=\left\langle n\right\vert \rho^{out}\left\vert n\right\rangle $ to define
the quality of the copies. As expected, the optimal fidelity for
$1\rightarrow2$ QPCCM is $F_{QPCCM}^{opt}=\frac{1}{2}+\frac{\sqrt{2}}{8}
=0.854$, which is higher than the optimal value $F_{QUCM}^{opt}=0.833$.

This quantum circuit of QPCCM is realized by using a two-qubit NMR quantum
computer, based on $^{13}C$ and the $^{1}H$ nuclei in Carbon-13 labelled
chloroform (Cambridge Isotopes) dissolved in d$_{6}$ acetone. The $^{13}C$
nucleus was used as qubit $a$, and $^{1}H$ as qubit $b$. The reduced
Hamiltonian of the 2-spin ensemble is given by $H=\omega_{a}I_{z}^{a}%
+\omega_{b}I_{z}^{b}+2\pi JI_{z}^{a}I_{z}^{b}$, where the first two terms
describe the free procession of spin $a$ ($^{13}C$) and $b$($ ^{1}H$) around
the static magnetic field with frequencies $100$Mhz and $400$ Mhz. $I_{z}%
^{a}\left(  I_{z}^{b}\right)  $ is the angular moment operator of $a $ ($b$)
in direction $\widehat{z}$, and the third term is the $J$ coupling of the two
spins with $J=214.5$Hz. $^{13}C$ nucleus's $T_{1}$\ relaxation time is $17.2$s
and it's $T_{2}$ relaxation time is $0.35$s. $^{1}H$ nucleus's $T_{1}$
relaxation time is $4.8$s and it's $T_{2}$ relaxation time is $3.3$s. In the
following, we describe how we experimentally realize the optimal
$1\rightarrow2$ QPCCM shown in Fig. 1.

(E1)\ Prepare the initial state: Initially the two qubits are in thermal
equilibrium with the environment and their state is described by the density
operator $\rho_{th}\propto\sigma_{z}^{a}+4\sigma_{z}^{b}$. We use the spatial
averaging technique~\cite{cory} to create the effective pure state
$|\uparrow\rangle_{a}\otimes|\uparrow\rangle_{b}$, or, in the density operator
form, $\mbox{$\textstyle\frac{1}{2}$}(1+\sigma_{z}^{a})\otimes\mbox
{$\textstyle\frac{1}{2}$}(1+\sigma_{z}^{b})$. The sequence of operations
leading to this state is shown in Fig. 2(a). We then perform a single hard
$\frac{\pi}{2}$ radio frequency ($rf$) pulse on $a$ qubit to generate one of
the desired equatorial state $\left\vert n\left(  \varphi\right)
\right\rangle _{a}^{in}=(\cos\varphi,\sin\varphi,0)$ with $ \varphi=\cos
(n\pi/12)$, $n=\{0,1,\cdots,23\}$.

\begin{figure}[ptb]
\includegraphics[width=\columnwidth]{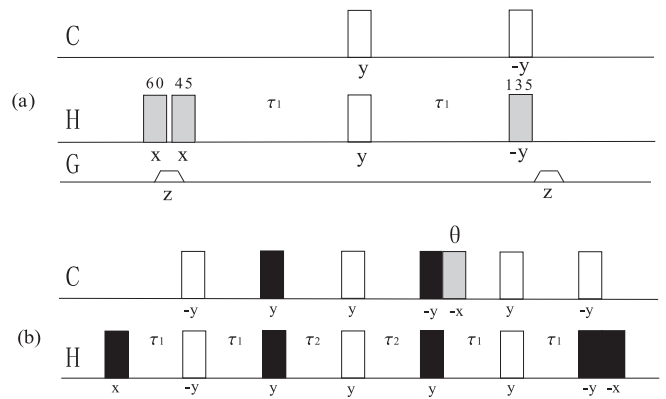}\caption{NMR pulse sequences.
The white and black boxes are $90^{o}$ and $180^{o}$\ pulses, while the grey
boxes are pulses with other flip angles shown above each box; pulse phases and
gradient directions are shown below each pulse. All $rf$ pulses are hard
pulses with regtangular shape and $5us$ pulse width. Delay times are $\tau
_{1}=1/(4J)$ and $\tau_{2} =\theta\cdot\tau_{1}/\pi$. In the QPCCM experiment
we take $\theta=\pi/4$ and change the value of $\theta\in\left[  0,2\pi\right]
$.}%
\end{figure}

(E2)\ Clone the input equatorial state: The quantum circuit of optimal
$1\rightarrow2$ QPCCM is described in Fig. 1 by fixing $\theta=\frac{\pi}{4}$.
This corresponds to a 2-qubit square root of SWAP gate. NMR pulse sequences
are developed by replacing this operation with an idealized sequence of NMR
pulses and delays. The resulting sequences are then simplified by combining
$rf$ pulses appropriately. Figure 2(b) shows the final pulse sequence to
demonstrate the optimal $1\rightarrow2$ QPCCM. All the $rf$ pulses are hard
pulses which hardly affect the state of $b$ qubit due to the heteronuclear
sample we used.

(E3) Measure and analyze: In principle, the quality of the copies, defined as
fidelity, can be calculated by $F=\left\langle n\right\vert \rho_{a\left(
b\right)  }^{out}\left\vert n\right\rangle $, where $\rho_{a\left(  b\right)
}^{out}$ is the reduced density matrix of a single qubit and can be obtained
from the density matrix $\rho_{ab}^{out}$. In NMR, one can use state
tomography technique to get $\rho_{ab}^{out}$ by applying a set of readout
pulses, but this has the disadvantage of requiring separate experiments.\ In
our experiment, we use a simpler method described in Ref.\cite{Cummins}: we
measure the two spectra of two output qubits individually, here the receiver
phase are set with the same phase as that of the input qubit measurement.
Therefore, the tracing out process can be implemented by integrating the
entire multiplet in each spectrum, comparing to the integration of the input
state spectrum, we can obtain the relative length of the output state vector
$r_{a(b)}^{\prime}$ in the same orientation as its input state vector, so the
fidelity between the input and output state can be calculated as
$F_{a(b)}=\frac{1}{2}(1+r_{a(b)}^{\prime}) $. Figure 3 shows the experimental
results from cloning the input equatorial state $\left\vert n\left(  0\right)
\right\rangle _{a}^{in}=(1,0,0) $. There are three spectra, corresponding to
the observable NMR signals of one input state and its two copies, that are
measured by setting the same receiver phase experimentally. The spectra do
have similar expected form (in-phase absorption signals at the outmost
positions of each multiplet).

\begin{figure}[ptb]
\includegraphics[width=\columnwidth]{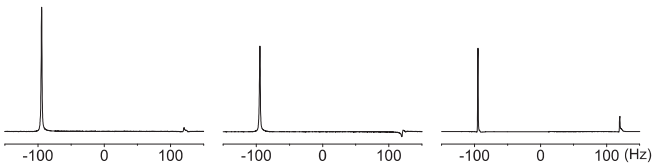}\caption{Experimental spectra of
cloning the input state $\left|  n\left(  0\right)  \right\rangle _{a}%
^{in}=(1,0,0)$. The left and middle spectra are the carbon spectra
corresponding to the input and output state of $a$ qubit, where the vertical
scales are in the same arbitrary units. The right spectrum is for the hydrogen
nucleus representing the output of $b$ qubit, where the vertical scale does
not share the same arbitrary units with those of the carbon spectra. From the
integration of the each multiplet, we obtain the fidelities of the two copies,
$F_{a}=0.842$ and $F_{b}=0.839$.}%
\end{figure}

An important feature of QPCCM is that all equatorial states are cloned equally
well and so it is necessary to study the behavior of pulse sequence when
applied to a wide range of states on equator. We have prepared a total of $24$
input equatorial states $\left\vert n\left(  \varphi\right)  \right\rangle
_{a}^{in}=(\cos\varphi,\sin\varphi,0)$ by changing the value $\varphi$\ with a
spacing of $15%
{{}^\circ}%
$ as shown in (E1). For each input state, we measure its spectrum and denote
it as a reference to calculate the quality of the two copies after the cloning
transformation described in (E2). Finally we measure each copy and calculate
the fidelity. Experimentally, we get the mean fidelity of this phase-covariant
cloning are $F_{a}=0.848\pm0.015$ for $a$ qubit and $F_{b}=0.844\pm0.015$ for
$b$ qubit, which are both close to the optimal theoretical value $0.854$.

Compared to the low fidelity of the NMR experiment for UQCM \cite{Cummins},
our near-optimal fidelity arises from the following reasons: (1) Less
decoherence effect -- the time used for cloning in our experiment is about
$5.3ms$, which is well within the decoherence time (about $350ms$ for $^{13}C$
nucleus and $3.3s$ for $^{1}H$ nucleus); while the time used for UQCM in Ref.
\cite{Cummins} was estimated about $400ms$, which is close to the decoherence
time with the value $720ms$ for two $^{1}H$ nuclei; (2) Simplicity -- it is
simpler to realize our economic 2-qubit QPCCM than to realize the 3-qubit UQCM
in Ref. \cite{Cummins}; (3) Pulses -- in our experiment all the $rf$ pulses
are hard pulses, which are more perfect than selective pulses, this is simply
achieved by using heteronuclear sample. In our experiments, small errors arise
as a result of the inhomogeneity of the static with $rf$ magnetic fields as
well as the variability of the measurement.

One important application of our efficient QPCCM is the quantum simulator of a
bit by eavesdropping on the four-state protocol of quantum
cryptography~\cite{NG,scarani}. (The UQCM plays the same role for the
six-state protocols \cite{bruss6states}). The four-state protocol (also known
as BB84) uses four quantum states, say $\left|  \pm\right\rangle _{x}=\frac
{1}{\sqrt{2}}\left(  \left|  0\right\rangle \pm\left|  1\right\rangle \right)
$ and $\left|  \pm\right\rangle _{y}=\frac{1}{\sqrt{2}}\left(  \left|
0\right\rangle \pm i\left|  1\right\rangle \right)  $, that constitute two
maximally conjugate bases. Alice chooses one of these four states in her qubit
(denoted by $a$) at random, then sends it to Bob. Whatever the state is, Eve
approximately clone two copies of this input state by inserting a
$1\rightarrow2$ QPCCM in the quantum channel. She then sends one copy to Bob
and stores another copy in her qubit $b$. Thereafter, Bob measures qubit $a$
in one of the two bases chosen at random. Finally Alice announces publicly the
basis she used for transmission of the signal, and in those cases in which Bob
measures in the same bases (these cases are useful, the others are discarded
by Alice and Bob). Eve, who now knows the bases Alice employed, measures $b$
qubit in order to estimate which signal Alice sent.

Experimentally, we realize the above process of eavesdropping attack as
following: (1) Prepare one of four BB84 states by\ using the same method that
we described in (E1); (2) Perform the quantum $1\rightarrow2$ QPCCM described
in Fig. 1. The pulse sequence to realize the network is shown in Fig. 2(b).
Note that $\theta$ is not fixed to $\pi/4$\ as in the previous optimal
$1\rightarrow2$ QPCCM realization, we set various rotation angles $\theta
\in\left[  0,2\pi\right]  $ of the pulse [marked as grey box in Fig.2(b)] with
a space of $\pi/12$; (3) For each experiment, we perform two measurements of
the NMR observable signals $\left\langle \sigma_{i}\right\rangle $ for Bob and
Eve individually, here $i\in\left[  x,y\right]  $ is the same base as that of
the input state. Recall another measurement for the input state which is used
as reference spectra, there are 3 measurements. Totally, we perform
$4\times24\times3=288$ experiments distinguished by $4$ BB84 states and $24$
rotation angles $\theta\in\left[  0,2\pi\right]  $ and $3$ measurements.
Theoreically, if Alice sends the state $\left|  \pm\right\rangle _{x}$, Bob
measures $\left\langle \sigma_{x}^{Bob}\right\rangle =\pm\frac{\cos\theta}{2}$
and Eve measures $\left\langle \sigma_{x}^{Eve}\right\rangle =\pm\frac
{\sin\theta}{2}$ in X-base; if Alice send state $\left|  \pm\right\rangle
_{y}$, Eve measures $\left\langle \sigma_{y}^{Eve}\right\rangle =\pm\frac
{\sin\theta}{2}$ while Bob measures $\left\langle \sigma_{y}^{Bob}%
\right\rangle =\pm\frac{\cos\theta}{2}$ in Y-base. Both the theoretical and
experimental results are plotted in Fig. 3, where the symmetry between
$\left|  +\right\rangle $ and $\left|  -\right\rangle $\ in each base is
clearly seen.

\begin{figure}[ptb]
\includegraphics[width=\columnwidth]{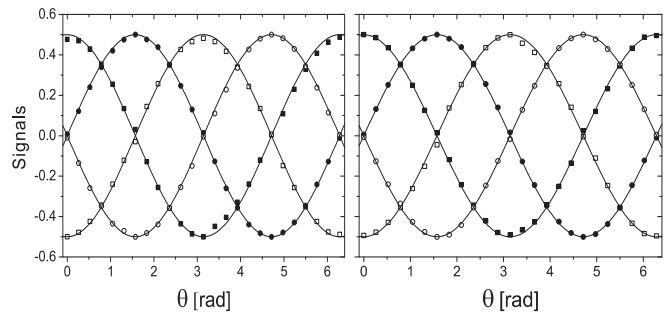}\caption{The normalized
observable NMR signals $\left\langle \sigma\right\rangle $\ versus the
rotation angle $\theta$\ of the phase-covariant cloning machine, for two
bases, $X$-base(at Left) and $Y$-base(at right). The lines correspond to
theoretical calculation. The filled and empty boxes (circles) correspond to
the experimental measurement from $a$ qubit ($b$ qubit), while the filled
(empty) circles and boxes correspond to the input state is $\left|
+\right\rangle $ ($\left|  -\right\rangle $).}%
\end{figure}

A main concern of the eavesdropping is to determine how much information an
eavesdropper can obtain from a given level of noise. For the above optimal
eavesdropping attack, regardless of the input BB84 state, Bob guesses
correctly the state sent by Alice with probability $F_{Bob}=\frac{1}%
{2}+\langle\sigma_{i}^{Bob}\rangle$ and makes an error $D_{Bob}=1-F_{Bob}%
=\frac{1}{2}-\langle\sigma_{i}^{Bob}\rangle$, where $i\in\{x,y\}$ is one
of\ the maximally conjugate bases; while Eve guesses correctly the state sent
by Alice with probability $F_{Eve}=\frac{1}{2}+\langle\sigma_{i}^{Eve}\rangle$
and makes an error $D_{Eve}=\frac{1}{2}-\langle\sigma_{i}^{Eve}\rangle$. As we
know, the mutual information is defined as $I=\frac{1}{2}+D\log_{2}D+\left(
1-D\right)  \log_{2}\left(  1-D\right)  $. From our experimental dates shown
in Fig. 4, we extract the Alice-Bob and Alice-Eve mutual information as a
function of the value of noise (QBER) defined as $QBER=\frac{1-\cos\theta}{2}%
$. Here $\theta\in\left[  0,\pi/2\right]  $ characterize the strength of Eve's
attack. The experimental results are shown in Fig. 5. We show the relation
between the mutual information and QBER, in agreement with the theoretical results.

\begin{figure}[ptb]
\includegraphics[width=0.8 \columnwidth]{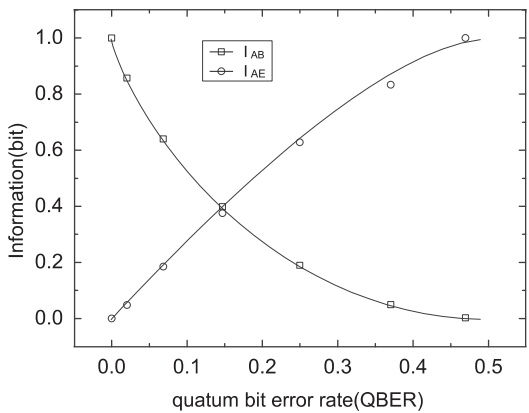}\caption{Mutual information
versus the quantum bit error rate. The lines correspond to theoretical
calculation. The boxes correspond to the experimental obtained mutual
information of Alice and Bob ($a$ qubit), while the circles correspond to the
experimental mutual information of Alice and Eve ($b$ qubit).}%
\end{figure}

In summary, we provide the first experimental demonstration of an efficient
and nearly optimal $1\rightarrow2$ QPCCM by using a 2-qubit NMR quantum
computer. Our approach cannot be extended to the UQC as it is known that a
3-qubit $1\rightarrow2$ UQC cannot be reduced to an efficient 2-qubit
network~\cite{thomas}. However, our efficient QPCCM has potential applications
as a simulator of eavesdropping techniques in quantum key distributions.

This project was supported by the National Nature Science Foundation of China
(Grants. No. 10075041 and No. 10075044) and Funded by the National Fundamental
Research Program (2001CB309300). We also thank supports from the ASTAR Grant
No. 012-104-0040 and Temasek Project in Quantum Information Technology (Grant
No. R-144-000-071-305). T.D. thanks supports from the Flemish Fund for
Scientific Research, the Inter-University Attraction Pole Program of the
Belgian government under grant V-18, the Concerted Research Action Photonics
in Computing, and the research council (OZR) of the VUB.

\end{document}